\documentclass[12pt]{article}
\usepackage{amsmath}
\usepackage{amsfonts}
\title{BRST approach to higher spin field theories}
\author{I.L. Buchbinder\footnote{joseph@tspu.edu.ru},
 V.A. Krykhtin\footnote{krykhtin@mph.phtd.tpu.edu.ru}}
\date{\it $^*$Department of Theoretical Physics Tomsk State Pedagogical
University \\ Tomsk, 634041, Russia
\\[2mm] $\dag$Tomsk Polytechnic University, Tomsk, 634050, Russia}

\begin{document}

\maketitle
\begin{abstract}
We develop the BRST approach to Lagrangian formulation for
massive bosonic and massless fermionic higher spin fields on a
flat space-time of arbitrary dimension. General procedure of
gauge invariant Lagrangian construction describing the dynamics
of the fields with any spin is given. No off-shell constraints
on the fields (like tracelessness) and the gauge parameters are
imposed. The procedure is based on construction of new
representations for the closed algebras generated by the
constraints defining irreducible representations of the
Poincare group. We also construct Lagrangians describing
propagation of all massive bosonic fields and massless fermionic
fields simultaneously.
\end{abstract}

\section{Introduction}

Construction of higher spin field theory is one of the fundamental problems of
high energy theoretical physics.
At present, there exist the various approaches
to this problem (see e.g. \cite{reviews} for reviews).
This paper is a brief review of recent development of BRST
approach to free higher spin field theory.
It is based on two papers \cite{0410215,0505092} devoted to
Lagrangian construction of free fermionic massless higher spin
fields and Lagrangian construction of free bosonic massive
higher spin fields respectively.
The main motivation for using BRST approach is to try to
construct the theory of interacting higher spin fields
analogously to string field theory.
The first natural step in constructing a higher spin interacting
model is formulation of the corresponding free theory.

The paper is organized as follows.
In sections \ref{ac-f} and \ref{ac-b}
we discuss operator algebras generated by primary
constraints which define irreducible representations of the
Pioncare group both in the massless fermionic and massive
bosonic cases respectively.
The sructure of the algebras proved to be the same and the
method of Lagrangian construction is explained in section
\ref{toysection} on the base of a toy model.
Then in sections \ref{Fermionic-L} and \ref{Bosonic-L} we applay
this method for Lagrangian construction both for massless
fermionic and massive bosonic fields respectively.
Section~\ref{Summary} summarizes the obtained results.

\section{Massless fermionic theory. Algebra of the constraints.}\label{ac-f}

It is well known that the totally symmetrical tensor-spinor field
$\Psi_{\mu_1\cdots\mu_n}$ (the Dirac index is suppressed),
describing the irreducible spin $s=n+1/2$ representation must
satisfy the following constraints (see e.g. \cite{BK})
\begin{eqnarray}
\gamma^\nu\partial_\nu \Phi_{\mu_1\cdots\mu_n}=0,
\label{irrep0}
&\qquad&
\gamma^\mu\Phi_{\mu\mu_2\cdots\mu_n}=0.
\end{eqnarray}
Here $\gamma^\mu$ are the Dirac matrices
$\{\gamma_\mu,\gamma_\nu\}=2\eta_{\mu\nu}$,
$\eta_{\mu\nu}=(+,-,\ldots,-)$.

In order to describe all higher tensor-spinor fields
together it is convenient to introduce Fock space
generated by creation and annihilation operators $a_\mu^+$,
$a_\mu$ with vector Lorentz index $\mu=0,1,2,\ldots,D-1$
satisfying the commutation relations
\begin{eqnarray}
\bigl[a_\mu,a_\nu^+\bigr]=-\eta_{\mu\nu}.
\label{c-a-comm}
\end{eqnarray}
These operators act on states in the
Fock space
\begin{eqnarray}
|\Phi\rangle&=&\sum_{n=0}^{\infty}\Phi_{\mu_1\cdots\mu_n}(x)
a^{+\mu_1}\cdots a^{+\mu_n}|0\rangle
\label{gstate}
\end{eqnarray}
which describe all half-integer spins simultaneously if the following
constraints are taken into account
\begin{eqnarray}
T_0|\Phi\rangle&=&0,
\qquad
T_1|\Phi\rangle=0,
\label{f-em-1}
\end{eqnarray}
where
\begin{eqnarray}
T_0=\gamma^\mu{}p_\mu,
\qquad
T_1=\gamma^\mu{}a_\mu,
\label{fconstr-1}
\end{eqnarray}
with $p_\mu=-i\frac{\partial}{\partial x^\mu}$.
If
constraints (\ref{f-em-1}) are fulfilled for the general state
(\ref{gstate}) then constraints (\ref{irrep0})
are fulfilled for each component $\Phi_{\mu_1\cdots\mu_n}(x)$ in
(\ref{gstate}) and hence the relations (\ref{f-em-1}) describe all
free higher spin fermionic fields together.
The constraints $T_0$, $T_1$ are primary constraints. They
generate all the constraints on the space of ket-vectors
(\ref{gstate}). Thus we get three more constraints
\begin{eqnarray}
&
L_0|\Phi\rangle=0,
\qquad
L_1|\Phi\rangle=0,
\qquad
L_2|\Phi\rangle=0,
\label{f-em-2}
\end{eqnarray}
where
\begin{eqnarray}
&
L_0=-p^2,
\qquad
L_1=a^\mu p_\mu,
\qquad
L_2={\textstyle\frac{1}{2}}\,a_\mu a^\mu.
\label{constraints}
\end{eqnarray}

Our purpose is to construct Lagrangian for the massless
fermionic higher spin fields on the base of BRST approach, therefore
we must construct Hermitian BRST operator.
In the case under consideration the constraints $T_0$, $L_0$ are
Hermitian, $T_0^+=T_0$, $L_0^+=L_0$, however the constraints
$T_1$, $L_1, L_2$ are not Hermitian.
Therefore we extend the set of the constraints adding three
new operators
\begin{eqnarray}
&
T_1^+=\gamma^\mu{}a_\mu^+,
\qquad
L_1^+=a^{+\mu} p_\mu,
\qquad
L_2^+={\textstyle\frac{1}{2}}\,a_\mu^+ a^{+\mu}
\label{bra-constraints}
\end{eqnarray}
to the initial constraints (\ref{fconstr-1}) and
(\ref{constraints}).
As a result, the set of operators $T_0$, $T_1$, $T_1^+$,
$L_0$, $L_1$, $L_2$, $L_1^+$, $L_2^+$
is invariant under Hermitian conjugation.
Taking hermitian conjugation of
(\ref{f-em-1}) and (\ref{f-em-2}) we see that the operators
(\ref{bra-constraints}) together with $T_0$ and $L_0$ are
constraints on the space of ket-vectors
\begin{eqnarray}
\langle\Phi|T_0=\langle\Phi|T_1^+=\langle\Phi|L_0=
\langle\Phi|L_1^+=\langle\Phi|L_2^+=0.
\end{eqnarray}

Algebra of operators
(\ref{fconstr-1}),
(\ref{constraints}), (\ref{bra-constraints})
is open in terms of commutators of these operators.
We will suggest the following procedure of consideration.
We want to use the BRST construction in the simplest (minimal)
form coresponding to closed algebras.
To get such an algebra we add to the above set of operators, all operators
generated by the commutators of
(\ref{fconstr-1}),
(\ref{constraints}), (\ref{bra-constraints}).
Doing such a way  we obtain one new operator
\begin{eqnarray}
G_0&=&-a_\mu^+a^\mu+{\textstyle\frac{D}{2}},
\end{eqnarray}
which arises from the commutators
\begin{eqnarray}
-{\textstyle\frac{1}{2}}[T_1,T_1^+]=
[L_2,L_2^+]=G_0,
\end{eqnarray}
and which is not a constraint neither in the space of
ket-vectors nor in the space of bra-vectors.
The resulting operators algebra may be found in \cite{0410215}.

Let us summarize what we have at the moment.
The structure of the operator algebra in the
fermionic case is as follows.
First we have hermitian operators $T_0$, $L_0$,
$G_0$. Two of them $T_0$ and $L_0$ are constraints both in the space
of ket-vectors and in the space of bra-vectors, another $G_0$ is not
a constrint neither in the space of ket-vectors nor in the space
of bra-vectors. Then we have pairs of mutually conjugated
operators ($T_1$, $T_1^+$), ($L_1$, $L_1^+$), ($L_2$, $L_2^+$).
One representative from the pairs is constraint in the space of
ket-vectors another representative is a constraint on the space
of bra-vectors.
The problem is to find BRST operator which reproduce equations
of motion (\ref{irrep0}) up to gauge transformations.

Let us turn to the massive bosonic case.

\section{Massive bosonic theory. Algebra of the constraints.}\label{ac-b}

It is well known that the totally symmetric tensor field
$\Phi_{\mu_1\cdots\mu_s}$, describing the irreducible spin-$s$
massive representation of the Poincare group must satisfy the
following constraints (see e.g. \cite{BK})
\begin{eqnarray}
&
(\partial^2+m^2)\Phi_{\mu_1\cdots\mu_s}=0,
\qquad
\partial^{\mu_1}\Phi_{\mu_1\mu_2\cdots\mu_s}=0,
\qquad
\eta^{\mu_1\mu_2}\Phi_{\mu_1\cdots\mu_s}=0.
\label{b-irrep0}
\end{eqnarray}

Analogously to the fermionic case,
in order to describe all higher integer spin fields
simultaneously we introduce Fock space
generated by creation and annihilation operators $a_\mu^+$,
$a_\mu$
satisfying the commutation relations (\ref{c-a-comm})
and define the operators
\begin{eqnarray}
\label{L0}
&L_0=-p^2+m^2,
\qquad
L_1=a^\mu p_\mu,
\qquad
L_2=\frac{1}{2}a^\mu a_\mu,
\label{b-constraints}
\end{eqnarray}
where
$p_\mu=-i\frac{\partial}{\partial{}x^\mu}$.
These operators act on states in the
Fock space
\begin{eqnarray}
|\Phi\rangle
&=&
\sum_{s=0}^{\infty}\Phi_{\mu_1\cdots\mu_s}(x)
a^{\mu_1+}\cdots a^{\mu_s+}|0\rangle
\label{b-gstate}
\end{eqnarray}
which describe all integer spin fields simultaneously if
the following constraints on the states take place
\begin{eqnarray}
&
L_0|\Phi\rangle=0,
\qquad
L_1|\Phi\rangle=0,
\qquad
L_2|\Phi\rangle=0.
\label{01}
\end{eqnarray}
If
constraints (\ref{01}) are fulfilled for the general state
(\ref{b-gstate}) then constraints (\ref{b-irrep0})
are fulfilled for each component $\Phi_{\mu_1\cdots\mu_s}(x)$ in
(\ref{b-gstate}) and hence the relations (\ref{01}) describe all
free massive higher spin bosonic fields simultaneously.

Constraints (\ref{b-constraints}) are all constraints in
the space of ket-vectors (\ref{b-gstate}).
Again, as in the fermionic case, in order to be possible to
construct hermitian BRST operator we must add to the constraints
(\ref{b-constraints}) their hermitian conjugated operators. Since
$L_0^+=L_0$ we add two operators
\begin{eqnarray}
&
L_1^+=a^{+\mu} p_\mu,
\qquad
L_2^+={\textstyle\frac{1}{2}}\,a_\mu^+ a^{+\mu}
\label{bra-b-constraints}
\end{eqnarray}
to the initial constraints
(\ref{b-constraints}).
As a result, the set of operators
$L_0$, $L_1$, $L_2$, $L_1^+$, $L_2^+$
is invariant under Hermitian conjugation.
Taking hermitian conjugation of
(\ref{01}) we see that the operators
(\ref{bra-b-constraints}) together with $L_0$ are
constraints in the space of ket-vectors
\begin{eqnarray}
\langle\Phi|L_0=
\langle\Phi|L_1^+=\langle\Phi|L_2^+=0.
\end{eqnarray}

Algebra of the constraints
(\ref{b-constraints}), (\ref{bra-b-constraints})
is not closed
and in order to construct BRST operator we must include in
the algebra
all the operators generated by
(\ref{b-constraints}), (\ref{bra-b-constraints}).
Thus we have to include in the algebra two more hermitian operator
\begin{eqnarray}
m^2
&\qquad\mbox{and}\qquad&
G_0=-a_\mu^+a^\mu+{\textstyle\frac{D}{2}}.
\label{mG}
\end{eqnarray}
which are obtained from the commutators
\begin{eqnarray}
[L_1,L_1^+]=L_0-m^2,
\qquad
[L_2,L_2^+]=G_0,
\end{eqnarray}
and which are not not constraints neither in the space of
ket-vectors nor in the space of bra-vectors.
The resulting operator algebra can be found in \cite{0505092}.

Let us summarize the structure of the operator algebra in the
bosonic case. It is the same as in the fermionic case.
First we have hermitian operators $T_0$, $L_0$, $m^2$,
$G_0$. Two of them $T_0$ and $L_0$ are constraints both in the space
of ket-vectors and in the space of bra-vectors, another $m^2$ and $G_0$
are not
constrints neither in the space of ket-vectors nor in the space
of bra-vectors. Then we have pairs of mutually conjugated
operators ($L_1$, $L_1^+$), ($L_2$, $L_2^+$).
One representative from the pairs is constraint in the space of
ket-vectors another representative is a constraint on the space
of bra-vectors.

In order to understand better the method used for construction
of BRST operator leading to the proper equations of motion
(\ref{irrep0}), (\ref{01}) it is useful to consider a toy model.

\section{A simlified model}\label{toysection}

Let us consider a model where the  'physical' states are defined by
the equations
\begin{eqnarray}
L_0|\Phi\rangle=0,
\qquad
L_1|\Phi\rangle=0,
\label{equations}
\end{eqnarray}
with some operators $L_0$ and $L_1$.
Let us also suppose that some scalar product
$\langle\Phi_1|\Phi_2\rangle$ is defined for the states $|\Phi\rangle$
and
let
$L_0$ be a Hermitian operator $(L_0)^+=L_0$ and let $L_1$ be
non-Hermitian  $(L_1)^+=L_1^+$ with respect to this scalar product.
In this section we show how to construct Lagrangian which will
reproduce (\ref{equations}) as equations of motion up to gauge
transformations.

In order to get the Lagrangian within BRST approach
we should begin with the Hermitian BRST operator.
However, the standard prescription does not allow to construct such a Hermitian operator
on the base of operators $L_0$ and $L_1$ if $L_1$ is non-Hermitian.
We assume to define the nilpotent Hermitian operator in the case under consideration as
follows.

Let us consider the algebra generated by the operators
$L_0$, $L_1$, $L_1^+$
and let this algebra takes the form
\begin{eqnarray}
\label{toyA1}
&&
[L_0,L_1]=[L_0,L_1^+]=0,
\\
&&
[L_1,L_1^+]=L_0+C,
\qquad
C=const\ne0.
\label{toyA2}
\end{eqnarray}

It is known (see e.g. \cite{0311257}) that in the case $C=0$
if we
construct Hermitian BRST operators as if all the operators
$L_0$, $L_1$, $L_1^+$ were the first class constraints
then this BRST operator
will reproduce the proper equations of motion
(\ref{equations}) up to gauge transformations.

Now
let us consider the case $C\ne0$.
In
this case the central charge $C$ plays the role analogous to
$m^2$ and $G_0$ in the algebras of the two previous sections.
If
we construct BRST operator as if the operators $L_0$, $L_1$,
$L_1^+$, $C$ are the first class constraints
we get a solution $|\Phi\rangle=0$ \cite{0505092}
what contradicts to (\ref{equations}).
This happens because we treat the operator $C$ as a constraint.

But the case $C=0$ may serve as a hint about solution to our
problem. Namely, we construct new representation of the algebra
(\ref{toyA1}), (\ref{toyA2}) with operator $C_{new}=0$ in this
representation.

Thus the solution is as follows.
We enlarge
the representation space of the operator algebra
(\ref{toyA1}), (\ref{toyA2}) by introducing the additional (new) creation and
annihilation operators and construct a new representation of the
algebra bringing into it an arbitrary parameter $h$.
The basic idea is to construct such a representation where the
new operator $C_{new}$  has the form
$C_{new}=C+h$.
Since parameter $h$ is arbitrary and $C$ is a central charge,
we can choose $h=-C$ and the
operator $C_{new}$ will be zero in the new representation.
After this we proceed as if operators
$L_{0new}$, $L_{1new}$, $L_{1new}^+$ are
the first class constraints.

For example, we can construct new representation of the operator
algebra (\ref{toyA1}), (\ref{toyA2}) as follows
\begin{align}
\label{toy0new}
&
L_{0new}=L_0,
&&
C_{new}=C+h,
\\
&
L_{1new}=L_1+hb,
&&
L_{1new}^+=L_1^++b^+.
\label{toy1new}
\end{align}
Here we have introduced the new bosonic creation and annihilation
operators $b^+$, $b$ with the standard commutation relations
\begin{math}
[b,b^+]=1.
\end{math}

In principle, we could set $h=-C$ and get $C_{new}=0$, but there
is one more equivalent scheme.
Namley we still consider $C_{new}$ as
nonzero operator including the arbitrary parameter $h$, but demand for state
vectors and gauge parameters to be independent on ghost $\eta_C$ as
before.
It can be shown \cite{0505092}
that these conditions reproduce that $h$ should be equal to $-C$.

Now if
we introduce the BRST construction
taking the operators in new representation
as if they were the first class constarints
\begin{eqnarray}
\label{toyQh}
Q_{h}
&=&
\eta_0L_0+\eta_CC_{new}
+\eta_1^+L_{1new}+\eta_1L_{1new}^+
-\eta_1^+\eta_1({\cal{}P}_0+{\cal{}P}_C),
\qquad
Q_{h}^2=0.
\end{eqnarray}
we shall get \cite{0505092} that equation $Q_h|\Psi\rangle=0$, where
\begin{eqnarray}
|\Psi\rangle
&=&
\sum_{k=0}^\infty
\sum_{k_i=0}^1
(\eta_0)^{k_1}
(\eta_1^+)^{k_2} ({\cal{}P}_1^+)^{k_3}
(b^+)^k|\Phi_{kk_1k_2k_3}\rangle
.
\label{stateh}
\end{eqnarray}
reproduces (\ref{equations}) up to gauge transformations.

Let us pay attention that operators $L_{1new}$ and $L_{1new}^{+}$ are not mutually
conjugate in the new representation if we use the usual rules for
Hermitian conjugation of the additional creation and
annihilation operators
$(b)^+=b^+$,
$(b^+)^+=b$.
To consider the operators $L_{1new}$, $L^{+}_{1new}$ as conjugate to each other we change a
definition of scalar product
for the state vectors (\ref{stateh})
\begin{math}
\label{toy-SP}
\langle\Psi_1|\Psi_2\rangle_{new}=
\langle\Psi_1|K_h|\Psi_2\rangle
,
\end{math}
with
\begin{eqnarray}
K_h&=&\sum_{n=0}^{\infty}|n\rangle\frac{h^n}{n!}\langle{}n|,
\qquad
\qquad
|n\rangle=(b^+)^n|0\rangle.
\end{eqnarray}
Now
the new operators $L_{1new}, L_{1new}^+$ are mutually conjugate
and the operator $Q_h$ is Hermitian
relatively the new scalar product (\ref{toy-SP})
since the following relations take place
\begin{eqnarray}
&&
K_hL_{1new}^+=(L_{1new})^{+}K_h,
\quad
K_hL_{1new}=(L_{1new}^{+})^+K_h,
\quad
Q_h^+K_h=K_hQ_h
.
\end{eqnarray}

Finally we note that the proper equations of motion
may be derived using the
following Lagrangian
\begin{eqnarray}
{\cal{}L}&=&\int d\eta_0 \langle\Psi|K_{-C}\Delta{}Q_{-C}|\Psi\rangle
\end{eqnarray}
where subscripts $-C$ means that we substitute $-C$ instead of
$h$. Here the integral is taken over Grassmann odd variable $\eta_{0}$.

\section{Lagrangians for massless fermionic fields}\label{Fermionic-L}

\subsection{New representation}
Let us first construct new representation for the operator
algebra.
Ones find
\begin{align}
L_{2new}^+&=
{\textstyle\frac{1}{2}}a_\mu a^\mu+b^+,
&L_{2new}&=
{\textstyle\frac{1}{2}}a_\mu^+ a^{+\mu}+
(b^+b+d^+d+h)b,\\
T_{1new}^+&=\gamma^\mu a_\mu+2b^+d+d^+,
&T_{1new}&=\gamma^\mu a_\mu^+-2(b^+b+h)d-d^+b,\\
G_{0new}&=-a_\mu^+a^\mu
+{\textstyle\frac{D}{2}}+2b^+b+d^+d+h,
\end{align}
with the other operators being unchanged.
Here $b^+$, $b$ are bosonic creation and annihilation operators
and $d^+$, $d$ are fermionic ones with the standard commutation
relations
$[b,b^+]=1$,
$\{d,d^+\}=1$.
Then we introduce the scalar product
in the Fock space so that
\begin{math}
\langle\Phi_1|\Phi_2\rangle_{new}
=
\langle\Phi_1|K|\Phi_2\rangle,
\end{math}
with operator $K$
\begin{eqnarray}
\label{K-f}
K&=&
\sum_{n=0}^\infty\frac{1}{n!}
  \Bigl(\,
     |n\rangle{}\langle{}n|\,C(n,h)
     -
     2d^+|n\rangle\langle{}n|d\,C(n+1,h)\,
  \,\Bigr),
\qquad
|n\rangle=(b^+)^n|0\rangle,
\\&&
C(n,h)=h(h+1)(h+2)\ldots(h+n-1),
\qquad
C(0,h)=1.
\label{C}
\end{eqnarray}
Now we construct BRST operator as if all the operators were the
first class constraints
\begin{eqnarray}
\nonumber
\tilde{Q}
&=&
q_0T_0
+
q_1^+T_{1new}+q_1T_{1new}^+
+\eta_0L_0+\eta_1^+L_1+\eta_1L_1^+
+\eta_2^+L_{2new}+\eta_2L_{2new}^+
\\&&
\nonumber
{}
+\eta_{G}G_{0new}
+i(\eta_1^+q_1-\eta_1q_1^+)p_0
-i(\eta_Gq_1+\eta_2q_1^+)p_1^+
+i(\eta_Gq_1^++\eta_2^+q_1)p_1
\\&&
\nonumber{}
+(q_0^2-\eta_1^+\eta_1){\cal{}P}_0
+(2q_1q_1^+-\eta_2^+\eta_2){\cal{}P}_G
+(\eta_G\eta_1^++\eta_2^+\eta_1-2q_0q_1^+){\cal{}P}_1
\\&&
{}
+(\eta_1\eta_G+\eta_1^+\eta_2-2q_0q_1){\cal{}P}_1^+
+2(\eta_G\eta_2^+-q_1^{+2}){\cal{}P}_2
+2(\eta_2\eta_G-q_1^2){\cal{}P}_2^+.
\label{f-auxBRST}
\end{eqnarray}
Let us notice that the BRST operator (\ref{f-auxBRST}) is
selfconjugate in the following sense
\begin{math}
\tilde{Q}^+K=K\tilde{Q},
\end{math}
with operator $K$ (\ref{K-f}).

\subsection{Lagrangians for the free fermionic fields of single
spin}

It can be shown \cite{0410215} that from equation
$\tilde{Q}|\Psi\rangle=0$ using gauge transformations we can
remove dependence of the fields and the gauge parameters on the
ghost fields $\eta_0$, ${\cal{}P}_0$, $q_0$, $p_0$ and
obtain equations of motion for field with given spin $s=n+1/2$
\begin{eqnarray}
&&
\Delta{}Q_{\pi}|\chi^{0}_{0}\rangle_n
+\frac{1}{2}\bigl\{\tilde{T}_0,\eta_1^+\eta_1\bigr\}
|\chi^{1}_{0}\rangle_n
=0,
\qquad\qquad
\tilde{T}_0|\chi^{0}_{0}\rangle_n
+
\Delta{}Q_{\pi}|\chi^{1}_{0}\rangle_n
=0.
\label{EofM-f}
\end{eqnarray}
Here $|\chi^{0}_{0}\rangle_n$ and $|\chi^{1}_{0}\rangle_n$ are
states with ghost numbers $0$ and $-1$ respectively
and
subscript $n$ indicates that the corresponding field
obeying
the condition
\begin{eqnarray}
\pi|\chi\rangle_n=\bigl(n+(D-4)/2\bigr)|\chi\rangle_n,
\label{f-pin}
\end{eqnarray}
with
\begin{eqnarray}
\pi&=&G_0+2b^+b+d^+d-iq_1p_1^++iq_1^+p_1
+\eta_1^+{\cal{}P}_1-\eta_1{\cal{}P}_1^+
+2\eta_2^+{\cal{}P}_2-2\eta_2{\cal{}P}_2^+.
\end{eqnarray}
Next $\tilde{T}_0=T_0-2q_1^+{\cal{}P}_1-2q_1{\cal{}P}_1^+$, $\{A,B\}=AB+BA$
and $Q_{\pi}$ is the part of $\tilde{Q}$ (\ref{f-auxBRST}) which
independent of $\eta_G$, ${\cal{}P}_G$, $\eta_0$, ${\cal{}P}_0$,
$q_0$, $p_0$
with substitution $h\to-\pi$ \cite{0410215}.

These field equations (\ref{EofM-f})
can be
deduced from the following Lagrangian
\begin{eqnarray}
{\cal{}L}_n
&=&
{}_n\langle\chi^{0}_{0}|K_\pi\tilde{T}_0|\chi^{0}_{0}\rangle_n
+
\frac{1}{2}\,{}_n\langle\chi^{1}_{0}|K_\pi\bigl\{
   \tilde{T}_0,\eta_1^+\eta_1\bigr\}|\chi^{1}_{0}\rangle_n
\nonumber
\\&&\qquad{}
+
{}_n\langle\chi^{0}_{0}|K_\pi\Delta{}Q_{\pi}|\chi^{1}_{0}\rangle_n
+
{}_n\langle\chi^{1}_{0}|K_\pi\Delta{}Q_{\pi}|\chi^{0}_{0}\rangle_n
,
\label{L1}
\end{eqnarray}
where the standard scalar product for the creation and
annihilation operators is assumed
and the operator $K_\pi$ is the operator $K$ (\ref{K-f}) where
the following substitution is done
$h\to-\pi$ \cite{0410215}.

The equations of motion (\ref{EofM-f}) and the
Lagrangian (\ref{L1}) are invariant under the gauge
transformations
\begin{eqnarray}
\delta|\chi^{0}_{0}\rangle_n
=
\Delta{}Q_{\pi}|\Lambda^{0}_{0}\rangle_n
 +
 \frac{1}{2}\bigl\{\tilde{T}_0,\eta_1^+\eta_1\bigr\}
 |\Lambda^{1}_{0}\rangle_n,
&\quad&
\delta|\chi^{1}_{0}\rangle_n
=
\tilde{T}_0|\Lambda^{0}_{0}\rangle_n
 +\Delta{}Q_{\pi}|\Lambda^{1}_{0}\rangle_n
,
\end{eqnarray}
which are reducible
\begin{align}
\delta|\Lambda^{(i)}{}^{0}_{0}\rangle_n
&=
\Delta{}Q_{\pi}|\Lambda^{(i+1)}{}^{0}_{0}\rangle_n
 +
 \frac{1}{2}\bigl\{\tilde{T}_0,\eta_1^+\eta_1\bigr\}
 |\Lambda^{(i+1)}{}^{1}_{0}\rangle_n,
&
|\Lambda^{(0)}{}^0_0\rangle_n=|\Lambda^0_0\rangle_n,
\label{GTi1}
\\
\delta|\Lambda^{(i)}{}^{1}_{0}\rangle_n
&=
\tilde{T}_0|\Lambda^{(i+1)}{}^{0}_{0}\rangle_n
 +\Delta{}Q_{\pi}|\Lambda^{(i+1)}{}^{1}_{0}\rangle_n,
&
|\Lambda^{(0)}{}^1_0\rangle_n=|\Lambda^1_0\rangle_n,
\label{GTi2}
\end{align}
with finite number of reducibility stages $i_{max}=n-1$ for spin
$s=n+1/2$.
It can be shown \cite{0410215} that the Lagrangian
(\ref{L1}) can be transformed to the Fang-Fronsdal Lagrangian
\cite{Fang} in four dimensions after eliminating the auxiliary
fields.

\subsection{Lagrangian for all half-integer spin fields}\label{LagrAll}

Now we turn to construction of Lagrangian describing propagation
of all half-integer spin fields simultaneously.
It can be show \cite{0410215} that it looks like
\begin{eqnarray}
{\cal{}L}&=&
\langle\chi^{0}_{0}|K_{\pi}\tilde{T}_0|\chi^{0}_{0}\rangle
+
\frac{1}{2}\langle\chi^{1}_{0}|K_{\pi}\bigl\{
   \tilde{T}_0,\eta_1^+\eta_1\bigr\}|\chi^{1}_{0}\rangle
\nonumber
\\
&&\qquad{}
+
\langle\chi^{0}_{0}|K_{\pi}\Delta{}Q_{\pi}|\chi^{1}_{0}\rangle
+
\langle\chi^{1}_{0}|K_{\pi}\Delta{}Q_{\pi}|\chi^{0}_{0}\rangle,
\label{Lall}
\end{eqnarray}
where
$|\chi^{0}_{0}\rangle$ and $|\chi^{1}_{0}\rangle$ are
states with ghost numbers $0$ and $-1$ respectively.
Then we have the following
gauge transformations for the fields
\begin{eqnarray}
\delta|\chi^{0}_{0}\rangle
=
\Delta{}Q_{\pi}|\Lambda^{0}_{0}\rangle
 +
 \frac{1}{2}\bigl\{\tilde{T}_0,\eta_1^+\eta_1\bigr\}
 |\Lambda^{1}_{0}\rangle,
&\qquad&
\delta|\chi^{1}_{0}\rangle
=
\tilde{T}_0|\Lambda^{0}_{0}\rangle
 +\Delta{}Q_{\pi}|\Lambda^{1}_{0}\rangle
 .
\end{eqnarray}
which are also reducible
\begin{align}
\delta|\Lambda^{(i)}{}^{0}_{0}\rangle
&=
\Delta{}Q_{\pi}|\Lambda^{(i+1)}{}^{0}_{0}\rangle
 +
 \frac{1}{2}\bigl\{\tilde{T}_0,\eta_1^+\eta_1\bigr\}
 |\Lambda^{(i+1)}{}^{1}_{0}\rangle,
&
|\Lambda^{(0)}{}^0_0\rangle=|\Lambda^0_0\rangle,
\label{GTi1all}
\\
\delta|\Lambda^{(i)}{}^{1}_{0}\rangle
&=
\tilde{T}_0|\Lambda^{(i+1)}{}^{0}_{0}\rangle
 +\Delta{}Q_{\pi}|\Lambda^{(i+1)}{}^{1}_{0}\rangle,
&
|\Lambda^{(0)}{}^1_0\rangle=|\Lambda^1_0\rangle.
\label{GTi2all}
\end{align}
Since the fields
$|\chi^{0}_{0}\rangle$ and $|\chi^{1}_{0}\rangle$
contain infinite number of spins and since the order of
reducibility grows with the spin value, then the order of
reducibility of the gauge symmetry will be infinite.

\section{Lagrangians for massive bosonic fields}\label{Bosonic-L}

\subsection{New representation for the algebra}\label{auxalg}

To construct new representation, we introduce two pairs of additional
bosonic annihilation and creation operators
$b_1$, $b_1^+$, $b_2$, $b_2^+$ with
the standard commutation relations
$[\,b_1,b_1^+]=
[\,b_2,b_2^+]=
1$
and construct new representation
as follows
\begin{align}
&
m^2_{new}=0,
&&
G_{0new}=-a_\mu^+a^\mu+{\textstyle\frac{D}{2}}
+b_1^+b_1+{\textstyle\frac{1}{2}}+2b_2^+b_2+h,
\\
&
L_{1new}^+=a^\mu p_\mu+mb_1^+,
&&
L_{1new}=a^{+\mu}p_\mu+mb_1,
\\
&
L_{2new}^+={\textstyle\frac{1}{2}}a^\mu a_\mu
 -{\textstyle\frac{1}{2}}b_1^{+2}+b_2^+,
&&
L_{2new}={\textstyle\frac{1}{2}}a^{+\mu}a_\mu
 -{\textstyle\frac{1}{2}}b_1^2+(b_2^+b_2+h)b_2,
\end{align}
with the other operators being unchanged.
Then
we change the definition of scalar product of vectors in the new
representation
\begin{math}
\langle\Phi_1|\Phi_2\rangle_{new}
=
\langle\Phi_1|K|\Phi_2\rangle,
\end{math}
with operator $K$ in the form
\begin{eqnarray}
\label{K}
K&=&\sum_{n=0}^\infty
     |n\rangle
     \frac{C(n,h)}{n!}
     \langle{}n|,
\qquad
\qquad
|n\rangle=(b_2^+)^n|0\rangle.
\end{eqnarray}
with $C(n,h)$ given in (\ref{C}).

Next we introduce the operator $\tilde{Q}$
as if all the operators were the first class constarints
\begin{eqnarray}
\tilde{Q}
&=&
\eta_0L_{0}+\eta_1^+L_{1new}+\eta_1L_{1new}^+
+\eta_2^+L_{2new}
+\eta_2L_{2new}^+
+\eta_{G}G_{0new}
-\eta_1^+\eta_1{\cal{}P}_0
-\eta_2^+\eta_2{\cal{}P}_G
\nonumber
\\&&
{}
+(\eta_G\eta_1^++\eta_2^+\eta_1){\cal{}P}_1
+(\eta_1\eta_G+\eta_1^+\eta_2){\cal{}P}_1^+
+2\eta_G\eta_2^+{\cal{}P}_2
+2\eta_2\eta_G{\cal{}P}_2^+
,
\label{auxBRST}
\end{eqnarray}

One can show that the operator (\ref{auxBRST})
satisfy the relation
\begin{math}
\tilde{Q}^+K=K\tilde{Q},
\end{math}
which means that this operator is Hermitian relatively the new scalar
product with operator $K$ (\ref{K}).

\subsection{Lagrangians for the massive bosonic field with given
spin}

It can be shown \cite{0505092} that we can construct Lagrangian for
the field with given spin as
\begin{eqnarray}
{\cal{}L}_n
&=&
\int d\eta_0\,\,
{}_n\langle\chi|K_{\sigma}Q_{\sigma}|\chi\rangle_n.
\label{Lagr-bos}
\end{eqnarray}
Here field $|\chi\rangle_n$ subject to the condition
\begin{eqnarray}
\sigma|\chi\rangle_n&=&(n+(D-6)/2)|\chi\rangle_n.
\label{chin}
\end{eqnarray}
with operator $\sigma$ being
\begin{eqnarray}
\sigma&=&G_0+b_1^+b_1
+2b_2^+b_2
+\eta_1^+{\cal{}P}_1-\eta_1{\cal{}P}_1^+
+2\eta_2^+{\cal{}P}_2-2\eta_2{\cal{}P}_2^+
.
\end{eqnarray}
Next $Q_\sigma$ is the part of operator $\tilde{Q}$
(\ref{auxBRST}) independent of the ghost fields $\eta_G$,
${\cal{}P}_G$ with the substitution $h\to-\sigma$.
Analogouly, operator $K_\sigma$ is operator (\ref{K}) where
substitution $h\to-\sigma$ be done.

The gauge symmetry induced by nilpotency of the operator
$Q_\sigma$ will be reducible with the first stage of
reducibility
\begin{align}
\label{GTb1}
\delta|\chi\rangle_n
&=Q_{\sigma}|\Lambda\rangle_n
&
gh(|\Lambda\rangle_n)=-1,
\\
\delta|\Lambda\rangle_n
&=Q_{\sigma}|\Omega\rangle_n,
&
gh(|\Omega\rangle_n)=-2
\label{GTb3}
.
\end{align}

\subsection{Unified description of all massive integer spin
fields}

It is evident, the fields with different spins $s=n$ may have different masses
which we denote $m_n$. First of all we introduce the state vectors with definite
spin and mass as follows
\begin{eqnarray}
|\chi,m\rangle_{n,m_{n}}&=&
|\chi\rangle_n\,\delta_{m,m_n},
\label{chim}
\end{eqnarray}
with $|\chi\rangle_n$ being defined in (\ref{chin})
and $m$ in (\ref{chim}) is now a new variable of the states $|\chi,m\rangle_{n,m_n}$.
Second, we introduce the mass operator $M$ acting on the variable $m$ so that
the states $|\chi,m\rangle_{n,m_n}$ are eigenvectors of the operator $M$ with
the eigenvalues $m_n$
\begin{eqnarray}
M|\chi,m\rangle_{n,m_{n}}&=&
m_n|\chi,m\rangle_{n,m_{n}}=
m|\chi,m\rangle_{n,m_{n}}
.
\label{M}
\end{eqnarray}

Construction of the Lagrangian decribing unified dynamics of fields with all
spins is realized
in terms of a single state $|\chi\rangle$ containing the fields
of all spins (\ref{chim})
\begin{eqnarray}
|\chi\rangle&=&\sum_{n=0}^\infty|\chi,m\rangle_{n,m_n}
\label{chi-All}
.
\end{eqnarray}

This Lagrangian describing a
propagation of all integer spin fields with different masses
simultaneously
looks like \cite{0505092}
\begin{eqnarray}
{\cal{}L}
&=&
\int d\eta_0\,\, \langle\chi|K_\sigma Q_{\sigma{}M}|\chi\rangle
.
\label{Lagr-All}
\end{eqnarray}

Let us turn to the gauge transformations.
Analogously to (\ref{chim}) we introduce the gauge parameters for the fields
with given spin and mass
\begin{eqnarray}
|\Lambda,m\rangle_{n,m_{n}}=|\Lambda\rangle_n\,\delta_{m,m_n}
,
&\qquad&
|\Omega,m\rangle_{n,m_{n}}=|\Omega\rangle_n\,\delta_{m,m_n}
\end{eqnarray}
and
analogously to (\ref{chi-All}) we denote
\begin{eqnarray}
|\Lambda\rangle=\sum_{n=0}^\infty|\Lambda,m\rangle_{n,m_n},
&\qquad&
|\Omega\rangle=\sum_{n=0}^\infty|\Omega,m\rangle_{n,m_n}
.
\end{eqnarray}
Summing up (\ref{GTb1}), (\ref{GTb3}) over all $n$
we find gauge transformation for the field $|\chi\rangle$
(\ref{chi-All})
and transformation for the gauge parameter $|\Lambda\rangle$
\begin{eqnarray}
\delta|\chi\rangle=Q_{\sigma{}M}|\Lambda\rangle,
&\qquad&
\delta|\Lambda\rangle=Q_{\sigma{}M}|\Omega\rangle.
\label{GT-All}
\end{eqnarray}

\section{Summary}\label{Summary}
We have developed the BRST approach to derivation of gauge
invariant Lagrangians both for
massless fermionic
and
massive bosonic
higher spin fields.
We investigated the (super)algebras generated by the constraints
which are necessary to define these irreducible
representations of the Poincare group and found that the
algebras have an identical structure.
In particular, the algebras contain operators which are not
constraints neither in the space of bra-vectors nor in the space
of ket-vectors.
For the operators which are not constraints to be made
harmless
this method includes construction of a new representation of the
algebra, after which the BRST operator can be obtained
as if all the operators were the first class constraints.

The main obtained results are
\begin{itemize}
\item
The Lagrangians for free arbitrary spin fields are constructed
in terms of completely symmetric tensor(-spinor) fields
(see eq.~(\ref{L1}) for massless fermionic fields
and
eq.~(\ref{Lagr-bos}) for massive bosonic fields)
in concise form.
No off-shell constraints (including tracelessness) on the fields
and the gauge parameters are used.
All the equations which define an irreducible representation of
the Poincare group (including tracelessness of the fields) are
consequences of the Lagrangian equations of the motion and the
gauge fixing.
\item
The models under consideration are reducible gauge theories.
In the bosonic case the models have the first order of reducibility
and in the fermionic case the order of reducibility grows with the value of
spin.
\item
Lagrangian describing propagation of all massless fermionic
fields simultaneously is constructed (\ref{Lall}).
Lagrangian describing propagation of all bosonic massive fields
(with different massess) simultaneously is constructed
(\ref{Lagr-All}).
\end{itemize}

There are several possibilities for extending our results.
This approach can be applied to Lagrangian construction
of fermionic massive fields and
to Lagrangian construction of higher spin fields (both
massive and massless)
with mixed symmetry of Lorentz indeces (see \cite{0101201} for
corresponding bosonic massless case).

\section*{Acknowledgements}
The work was supported in part by
the INTAS grant, project
INTAS-03-51-6346,
the RFBR grant, project No.\ 03-02-16193,
the joint RFBR-DFG grant, project No.\ 02-02-04002,
the DFG grant, project No.\ 436 RUS 113/669,
the grant for LRSS, project No.\ 1252.2003.2.

\end{document}